\documentclass{sigchi}

\CopyrightYear{2019}
\setcopyright{rightsretained}

\toappear{
Permission to make digital or hard copies of all or part of this work
for personal or classroom use is granted without fee provided that
copies are not made or distributed for profit or commercial advantage
and that copies bear this notice and the full citation on the first
page. Copyrights for third-party components of this work
must be honored. For all other uses, contact the owner/author(s).
}

\usepackage{balance}       
\usepackage{graphics}      
\usepackage[T1]{fontenc}   
\usepackage{txfonts}
\usepackage{mathptmx}
\usepackage[pdflang={en-US},pdftex]{hyperref}
\usepackage{color}
\usepackage{booktabs}
\usepackage{textcomp}

\usepackage{microtype}        
\usepackage{ccicons}          

\usepackage{todonotes}

\def\plaintitle{Warning Signs in Communicating the Machine Learning Detection Results of Misinformation with Individuals}

\def\emptyauthor{}
\def\plainkeywords{Fake news; warning sign; machine learning; political news.}

\makeatletter
\def\url@leostyle{%
  \@ifundefined{selectfont}{
    \def\UrlFont{\sf}
  }{
    \def\UrlFont{\small\bf\ttfamily}
  }}
\makeatother
\def\pprw{8.5in}
\def\pprh{11in}

\definecolor{linkColor}{RGB}{6,125,233}
\hypersetup{%
  pdftitle={\plaintitle},
  pdfauthor={\emptyauthor},
  pdfkeywords={\plainkeywords},
  pdfdisplaydoctitle=true, 
  bookmarksnumbered,
  pdfstartview={FitH},
  colorlinks,
  citecolor=black,
  filecolor=black,
  linkcolor=black,
  urlcolor=linkColor,
  breaklinks=true,
  hypertexnames=false
}

\begin{document}

\title{\plaintitle}

\numberofauthors{1}
\author{%
  \alignauthor{Limeng Cui\\
    \affaddr{The Pennsylvania State University}\\
    \affaddr{University Park, PA}\\
    \email{lzc334@psu.edu}}\\
}

\maketitle

\begin{abstract}
  With the prevalence of misinformation online, researchers have focused on developing various machine learning algorithms to detect fake news. However, users' perception of machine learning outcomes and related behaviors have been widely ignored. Hence, this paper proposed to bridge this gap by studying how to pass the detection results of machine learning to the users, and aid their decisions in handling misinformation. An online experiment was conducted, to evaluate the effect of the proposed machine learning warning sign against a control condition. We examined participants' detection and sharing of news. The data showed that warning sign's effects on participants' trust toward the fake news were not significant. However, we found that people's uncertainty about the authenticity of the news dropped with the presence of the machine learning warning sign. We also found that social media experience had effects on users' trust toward the fake news, and age and social media experience had effects on users' sharing decision. Therefore, the results indicate that there are many factors worth studying that affect people's trust in the news. Moreover, the warning sign in communicating machine learning detection results is different from ordinary warnings and needs more detailed research and design. These findings hold important implications for the design of machine learning warnings.
\end{abstract}

\category{H.5.2.}{Information interfaces and presentation}{User Interfaces}{}


\section{Introduction}

Fake news is a type of false information to deliberately mislead or manipulate public opinion, through traditional mass media and recent online social media. In recent years, social media platforms (e.g., Facebook, Twitter) have made it possible for individuals to produce, consume, and share different information. A report on the 2016 election indicates that fake news websites rely on online social media for 48\% of traffic, which is a much higher share than of other sources \cite{allcott2017social}. With the blurring of the boundaries between information sources and recipients, it is difficult to control the quality of the information that people are exposed to. Especially, it must be acknowledged that people are not necessarily good at evaluating the quality of online information.

With more fact-checking work being done by machine learning algorithms \cite{shu2017fake}, this paper studied how to communicate the detection results with users and help them make decisions subsequently. Since Twitter has become the main source of news \cite{broersma2013twitter}, this project focuses on news and user's behaviors on Twitter. Two research questions are proposed:

RQ1: Will a machine learning warning help users judge the authenticity of news compared with the condition in which there is no warning?

RQ2: Will a machine learning warning influence user's subsequent behaviors such as clicking or sharing news compared with the condition in which there is no warning?

This paper investigated the two research questions by assessing the relationship between the presence of a warning sign and people's trust in online news.

\section{Literature Review}
A specific warning, which gives details about the continued influence of misinformation, succeeded in reducing the continued reliance on outdated information \cite{ecker2010explicit}. Clayton et al. further suggested that although the exposure to a general warning did not affect the perceived accuracy of headlines, it decreased the individual's belief in the accuracy \cite{clayton2019real}. Some researchers raised the opposite view that alerted individuals may perform worse \cite{szpitalak2010warning}. Pennycook et al. claimed that the exposure increases subsequent perceptions of accuracy, and tagging such stories as disputed is not effective \cite{pennycook2018prior}. The two venues of claims motivate this project.

\subsection{Warning Sign}
Banner or pop-up warnings are often seen when there is a potential threat \cite{zhang2014effects}. These warning signs can release a perceived risk from the information from social media, remind people of suspicious or false information. A logical result would be a decrease in users' trust toward the news with the presence of a warning sign. So we propose that:

H1: With the presence of a warning sign, participants will have less trust in the information, as well as a lower tendency for sharing the information.

\section{Methods}
The between-subjects online study was conducted to evaluate the effects of machine learning warnings in conveying the fake news detection results. In addition to the warning condition, a control condition, in which no warning is presented, is also conducted. Participants were asked to make detection and sharing decisions on fake and real news.

\subsection{Pilot Study}
To understand the issues that participants are concerned with and the way they think about them, an initial interview with 4 individuals (2 female, 2 male) was conducted. Participants were students from The Pennsylvania State University, with a variety of majors. Through 10 minute face-to-face interviews, they were asked following open-ended questions:

Q1: Have you seen fake news on SNSs, such as Twitter, Facebook? How do you judge the authenticity of the news? What factors will affect your judgment?

Q2: When you judge the authenticity of the news, will you rely on the warning information embedded on the website?

Q3: What do you do when you find out that a piece of news is fake? Will you still comment or retweet it?

Before the interview started, I asked the participants for permission to record. Participants were compensated with \$5 Starbucks gift cards.

Participants reported varying levels of enthusiasm to browse news on the internet and SNSs. Some participants know very well about the news, while others do not care. This makes them very different in their ability to identify real and fake news. For example, By looking at the news in Figure \ref{fig:design}, Individual B said: \textit{``I would think that is real.''}, while Individual A said: \textit{``I don't think it's true. [\dots] Just based on instinct like, `donating \$1 billion' is too generous to be real.''}

In addition to using common sense and prior knowledge to judge the authenticity of the news, some participants also mentioned that the source of the news is an important basis for their judgment of the real and fake news:

\textit{``I look at the photo, I look at the source, and I look at where I found it. For example, if I'm reading from [\dots], I'd be very disinclined to think that's fake. Because it's a reputable news source.''} (Individual T)

The same applies to the warning sign. After carefully asking about the design of the warning sign, Individual S said: \textit{``If this sign comes from a dependable or trustable source, then I would believe it and I would be cautious about that content. But if I don't know the source of that warning sign, I will definitely be more vulnerable even if there is a warning sign.''}

Participants showed a very different attitude towards the warning sign. When asked if he would trust warning sign, Individual B answered: \textit{``Yeah, definitely. I just whenever I see things I just kind of assume it's real.''} Individual A had the same opinion: \textit{``If I see it (the news in Figure \ref{fig:design}), I would probably click on it to see if it is true. But if I was suspicious already and that (the warning sign) was there I would think that it (the news) is not true.''}

All participants said they would not comment or share fake news. They skip it and also not report it.

It is worth noting that political tendency also plays an important role when people are facing online news:

\textit{``It depends on which source I believe.''} (Individual T)

Through the interview, I found individual's frequency of reading news online and interest in political news may be related to the judgment of news. Thus these two points were included in the survey study.

\subsection{Survey Study}
Based on the results of the interview, we had a general understanding of users' perceptions of warning signs, related to their judgment of news authenticity and willingness to share. In this section, the between-subjects online study investigating the effect of a machine-learning warning sign in mitigating fake news was conducted. In addition to the warning conditions, a control group in which no warning was presented, was also included in the study.

\subsubsection{Participants}

The study was conducted on Amazon Mechanical Turk (MTurk). 100 MTurk workers were recruited on March 25, 2019. All participants were (1) at least 18 years old; (2) located at the United States; and (3) with a human intelligence task (HIT) approval rate above 95\%. Participants were allowed to participate in the study once.

\subsubsection{Materials}

\begin{figure}[!htb]
    \centering
    \includegraphics[width=0.45\textwidth]{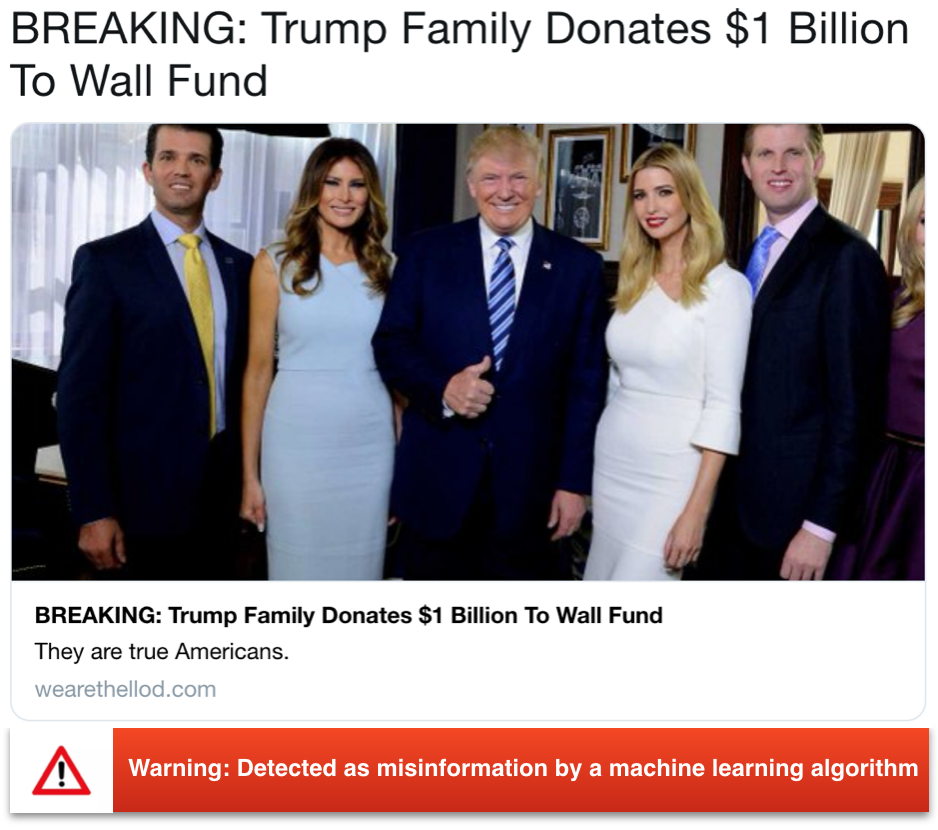}
    \caption{Warning sign design. A piece of fake news is at the top. A warning sign which indicates that the above news is disputed by a machine learning algorithm is at the bottom.}
    \label{fig:design}
\end{figure}

20 news headlines were created in the format of Twitter, consisting of a picture, source, header, and a short description (see Figure \ref{fig:design}). 10 verified fake news headlines, and 10 verified real news headlines were chosen from ``politifact.com'', which is a well-known third-party fact-checking website. Proposed machine learning warning is attached to the bottom of the fake news. Figure \ref{fig:design} gives a depiction of the warning design.

The selected news were released from January to March in 2019, and the topic of news was limited to politics because political news is one type of the most popular news that most individuals read every day, so most of the people have the certain sense to judge its credibility without professional knowledge.

\subsubsection{Procedure}

Participants completed a demographic questionnaire
that asked for age, gender. Then we asked participants completed additional questions about their social media experience, interest in politics, factors that impact their decisions.

20 news headlines with and without warnings were shown to a participant along with the questions, respectively. 20 news were presented one at a time in a randomized order. The participants were asked to judge the accuracy and decide their willingness to share the news on a 5-point Likert scale, respectively (1 means ``Very inaccurate'' or ``I would never share news like this one'', 5 means ``Very accurate'' or ``I would love to share news like this one''). Each participant was compensated for \$0.5 for the completion of the task.

\subsubsection{Measures}

Among the 100 MTurk workers, there were 68 male, 30 female. 2 people chose not to disclose. Participants came from different age groups, with 18.0\% between 18 to 25 years, 35.0\% between 26 to 30 years, 20.0\% between 31 to 35 years, 11.0\% between 36 to 45 years, 11.0\% between 46 to 55 years and 5.0\% above 55 years.

The social media experience was ranked into five degrees from 1 to 5 based on the frequency of browsing the web in a week: ``Extremely likely (Everyday)'', ``Very likely (Several times a week)'', ``Moderately likely (Once or twice a week)'', ``Slightly likely (Less often)'' and ``Not at all likely (Never)''. Most workers reported their social media experience on either ``Extremely likely'' (35.0\%) or ``Very likely'' (37.0\%); 20\% workers reported on ``Moderately likely''; and only 6.0\% and 2.0\% workers reported on ``Slightly likely'' and ``Not at all likely'' ($M=2.03, SD=.99$).

Similarly, the interest in politics was categorized into five types from 1 to 5: ``Extremely interested'', ``Very interested'', ``Fairly interested'', ``Not very interested'' and ``Not at all interested''. Most workers reported their social media experience on either ``Very interested'' (37.0\%); 29\% and 21\% workers reported on  ``Fairly interested'' and ``Extremely interested'' respectively; and only 9\% and 4\% workers were ``Not very interested'' and ``Not at all interested'' respectively ($M=2.38, SD=1.04$).

The demographic distributions were similar among the two conditions.

\begin{figure}[!htb]
    \centering
    \includegraphics[width=0.45\textwidth]{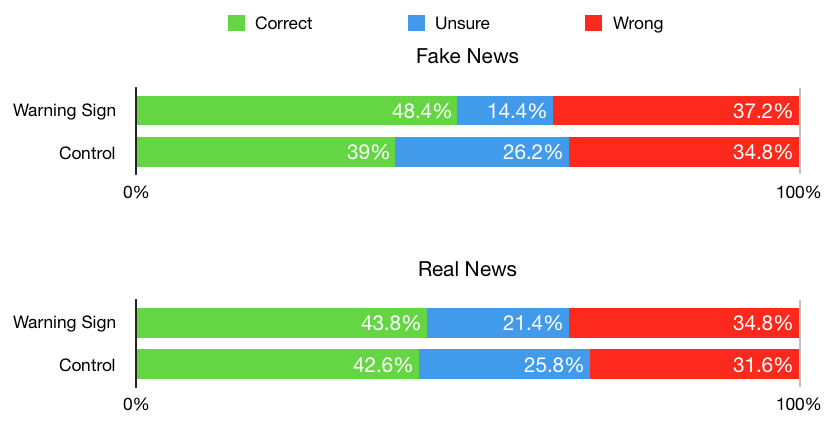}
    \caption{Participants' judgment toward fake news and real news.}
    \label{fig:judgment}
\end{figure}

Participants' trust toward the news were assessed with the aforementioned 5 points Likert scale including ``Very inaccurate'' as 1, ``Inaccurate'' as 2, etc. For fake news, participants gave lower scores with the presence of a warning sign ($M=2.88, SD=.84$) in contrast to the control condition ($M=2.70, SD=1.18$). Interestingly, with the presence of a warning sign, participants also gave lower scores to real news ($M=3.07, SD=.84$) compared with the control condition ($M=3.12, SD=.78$). Participants' scoring results were further grouped into three categories according to the ground truth: ``Correct'', ``Unsure'' and ``Wrong''. The distribution of participants' judgment toward fake news and real news is shown in Figure \ref{fig:judgment}. After introducing the warning sign, participants' uncertainty about the authenticity of the news decreased, with 26.2\% to 24.4\% toward the fake news. Interestingly, the uncertainty toward the real news was also dropped from 25.8\% to 21.4\%, where there was no warning sign attached to the real news in both conditions. In addition, with the presence of a warning sign, participants' detection accuracy of fake news increased from 39\% to 48.4\% compared with the control group.

Similar patterns could be observed from participants' sharing decision. With the presence of warning sign, participants were less willing to share both the fake news ($M=2.33, SD=1.28$) and real news ($M=2.53, SD=1.18$) in contrast to the control conditions (fake news: $M=2.79, SD=1.30$; real news: $M=2.76, SD=1.20$).

\section{Data Analysis}
\subsection{Effects of Machine Learning Warnings}
In order to test the effects of the machine learning warning sign on users' detections of the news and their sharing behaviors, a series of analyses of covariance (ANCOVA) were conducted, controlling for age, gender, and other demographic information. Warning sign's effects on participants' trust toward the fake news were not significant. We could not find enough statistical evidence to support that the warning sign can lower individuals' trust toward the fake news in contrast to the condition without the attached warning sign. However, results showed significant effects for the social media experience on participants' trust toward the fake news, $F=4.009, p<.05$. More experience on social media indicates less trust in fake news. Thus, for RQ1, there is no sufficient evidence to support that a warning sign could help users judge the authenticity of the news.

ANCOVA also indicated that age and social media experience had significant effects on users' sharing decision, $F=6.221, p<.05$ and $F=4.592, p<.05$. However, no statistical evidence supports the warning sign's effects on participants' sharing decision toward the fake news, which answered RQ2.

In sum, from the data collected, the warning sign would not impact users' trust and sharing decision.

\subsection{Modeling the Effects of Machine Learning Warnings}

\begin{figure}[!htb]
    \centering
    \includegraphics[width=0.45\textwidth]{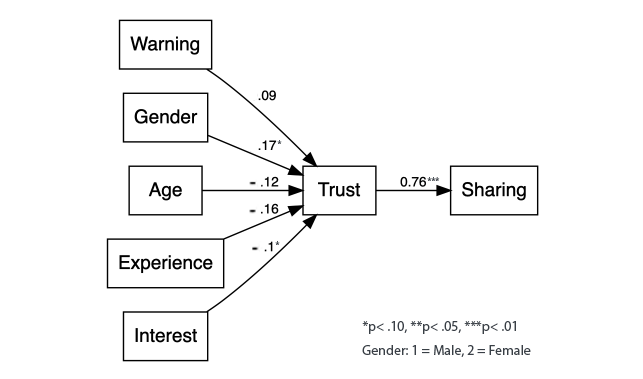}
    \caption{SEM predicting users' trust and sharing decision}
    \label{fig:sem}
\end{figure}

The relationships between the warning sign and other variables were tested using a structural equation model (SEM) shown in Figure \ref{fig:sem}, which yielded a good fit $\chi^2=42.74339, df=15, p < .001$.

The findings suggest the presence of machine learning warning sign could not reduce users' trust and sharing decision toward the fake news. Interestingly, users' sharing behavior is related to their trust. Specifically, users tend to share what they think is accurate.

\section{Discussion and Conclusions}
Through data analysis, the impacts of the proposed machine learning warning sign were examined. There is no statistical evidence to support that the proposed warning sign can help people better detect fake news and decrease their willingness to share fake news. However, results showed significant effects for the social media experience on people's trust toward the fake news. Also, age and social media experience had significant effects on people's sharing decision.

These results might go against many people's intuitive notion that a warning sign should lower people's trust, and raise many questions about factors that build people's trust, and factors that might influence people's decisions about whether to comment on or share particular news. If a warning sign---as a common and intuitive way to communicate with users---can not let users believe the potential misinformation in the news, what other factors can? Alternatively, can we find other better design to pass the detection results to users? This study provided an open-ended question. As a result, we need further studies to inform us what factors are of major influence on people's trust and sharing decision.

The results can be informative for researchers. Because researchers usually concentrate on how to detect the misinformation more accurately. Our results show that continuously increasing the accuracy might not be of much help in aiding people's decision in handling misinformation, as the machine learning warnings we used were consistent with ground truth. So, researchers should look for other ways to better communicating the machine learning detection results of misinformation.

This study can also provide some design implications for designers. An important implication for warning sign design is that it has to be designed carefully and strategically in triggering appropriate cognitive heuristics. While warning signs may be appropriate for garbing users' attention, designers may want to rethink this strategy for false information. Follow-up studies could focus on other mechanisms for persuasive appeals are linked to machine learning warning signs, as well as their effects on users' trust and sharing decision.

\balance{}

\bibliographystyle{SIGCHI-Reference-Format}
\bibliography{sample}

\end{document}